# Ethics and Artificial Intelligence Adoption

Martim Veiga
ISEG, Universidade de Lisboa
Lisboa, Portugal
l53080@aln.iseg.ulisboa.pt

Carlos J. Costa
ISEG, Universidade de Lisboa
Lisboa, Portugal
cjcosta@iseg.ulisboa.pt

*Abstract*—In recent years, we have witnessed a marked development and growth in Artificial Intelligence. The growth of the data volume generated by sensors and machines, combined with the information flow resulting from the user actions on the Internet, with high investments of the governments and the companies in this area, provided the practice and developed the algorithms of the Artificial Intelligence However, the people, in general, started to feel a particular fear regarding the security and privacy of their data and the theme of the Artificial Intelligence Ethics began to be discussed more regularly. The investigation aim of this work is to understand the possibility of adopting Artificial Intelligence nowadays in our society, having, as a mandatory assumption, Ethics and respect towards data and people's privacy. With that purpose in mind, a model has been created, mainly supported by the theories that were used to create the model. The suggested model has been tested and validated through Structural equation modeling based on data taken back from the respondents' answers to the questionnaire online: 237 answers, mainly from the Investigation Technologies area. The results obtained enabled the validation of seven of the nine investigation hypotheses of the proposed model. It was impossible to confirm any association between the Social Influence construct and the variables of Behavioral Intention and the Use of Artificial Intelligence. The aim of this work was accomplished once the investigation theme was validated and proved that it is possible to adopt Artificial Intelligence in our society, using the Attitude Towards Ethical Behavioral construct as the mainstay of the model.

## I. Introduction

We live in a world filled with the impact of technology on people, companies, and society in general, in the form of machines, sensors, cameras, or algorithms, which work based on data generated daily. Building artificial intelligence algorithms could be one of the most incredible feats in human history, both for its inherent difficulty and for the impact this feat would have on society's short-term, medium-term, and long-term future. On the other hand, Artificial Intelligence could be pretty dangerous in the future of Humanity, even admitting the possibility of destroying it if it is not adequately controlled [1].

The massive flow of information and connections between people generates a relatively high volume of data, contributing to the creation and training of Artificial Intelligence algorithms in Machine and Deep Learning. The practically exponential growth can be explained by the increase in the volume of data, which allows for training more precisely in the machines and algorithms of Artificial Intelligence, by the high amounts invested by research and development companies, especially Chinese and American, and by the tax incentives of the governments of certain countries.[1] [2]

Currently, there is no consensus regarding the advantages and the need to implement algorithms and tools based on artificial intelligence in society. Although many people are aware of the added value that the excellent use of this technology can bring to their personal and professional lives, another group is more skeptical about the dangers that it can bring to you and your data.

The scientific literature has focused on ethics in artificial intelligence, identifying several concerns. There is some worry associated with the area of Artificial Intelligence, and the most relevant fears are related to the fear of certain people seeing the privacy and security of their data affected and the fear of seeing their professions occupied by robots capable of performing the same tasks with greater precision and speed.

In order to answer the research question presented, a Literature Review will initially be carried out, which will cover some of the most relevant concepts of the theme. Then, the proposed model will be elaborated and presented, resulting from the literature review's main conclusions, the research hypotheses, and the questionnaire structure used to validate the model. The SEM approach was used to validate the model. Finally, the main conclusions of the work and possible future work will be presented.

## II. Literature review

Like many technologies, the concept of Artificial Intelligence (AI) has struggled to generate consensus within the expert community. While there is no universally accepted definition, certain theoretical principles and foundational values provide a shared framework for understanding. Technically, AI is often described as a sophisticated algorithm designed to process lower-dimensional algorithms. In this context, an algorithm refers to instructions processed sequentially, enabling a machine to perform tasks or follow orders as specified.[3]

The origins of Artificial Intelligence trace back to 1956 when Marvin Minsky and John McCarthy organized an eight-week workshop known as the Dartmouth Summer Research Project on Artificial Intelligence (DSRPAI) at Dartmouth College in New Hampshire. This landmark event aimed to explore the creation of machines capable of simulating human intelligence. Promising advancements in AI research marked the years following this workshop. However, political and economic factors soon stifled progress. In 1973, the United States Congress criticized the expenditures on AI research, while English mathematician James Lighthill published a report for the British Research and Science Institute, questioning the field's viability. In response, the U.S. and U.K. governments significantly reduced funding for AI research, leading to a period known as the "AI Winter"- AI research experienced a resurgence in the late 20th century. In 1997, IBM's Deep Blue chess program achieved a significant milestone by defeating world champion Garry Kasparov. Deep Blue processed 200 million possible moves per second to determine the optimal strategy, marking a significant achievement in computational intelligence. In 2012, Google advanced AI further by creating one of the first deep learning algorithms capable of recognizing images of cats within a dataset. This algorithm used a neural network with over 16,000 processors and one billion connections trained on random YouTube video frames. The system autonomously defined a "cat" based on the provided images. [1]

Based on the concepts of Diffusion and Innovation, Rogers' Theory, or Theory of the Diffusion of Innovation, was published in 1995. Although it can be applied in several areas, this theory is regularly used in adopting and implementing information technologies within communities with specific characteristics.[4]

According to Rogers's theory, the social system influences the diffusion of innovation and is defined as the necessary correlation of parts to solve everyday needs through problem-solving. The structure of the Social System leads to an awareness of the individuals in the group in question. Third parties' social influence affects an innovation's adoption rate. [4]

The adoption and acceptance of information technology and its consequent use derive from the perception of its usefulness in personal performance and its expected difficulty [5]. To understand (the way) how users would react, as well as the factors that influence them to use a particular new technology.

The Perception of Ease of Use (PFU) is defined as the degree of difficulty the user will feel when using a particular technology. On the other hand, the Perception of Utility (PU) can be characterized as the benefit of adopting technology to the person and their performance. The analysis of the two previous dimensions leads to the creation of an opinion about the technology that will result in an Attitude (A) and, later, in the Behavioral Intention (CI) of the adoption of this system. [4]

The Perception of Ease of Use and the Perception of Utility resulted in the creation of the Self-Efficacy Theory. The judgment of self-efficacy can be described as the belief in the ability of someone to perform a set of tasks necessary in a given situation, while the judgment of the result is related to the achievement of value when performing the task. Briefly, the perception of ease of use corresponds to self-efficacy, and the perception of usefulness is equivalent to the judgment of the outcome [5].

The Cost-Benefit Paradigm, derived from the Behavioral Decision Theory, supports the question of the usefulness and ease of use of information technology. People's decisions vary according to trading strategies, in which the effort and adjacent benefits are measured, with the cost representing the effort and the benefit referring to the results achieved. It is then observed that the distinction between effort and the resulting performance is equivalent to the distinction between the perception of ease of use and the perception of utility [5].

Adopting an innovation by a person has decision-making variables such as complexity, competitive advantage, and compatibility. Complexity is the perception of difficulty associated with using a given product [4], a concept quite similar to the Ease of Use dimension proposed in the TAM model. Ease of Use and Effectiveness are the two most relevant factors for the end user). Once again, the variables of the TAM model are supported by studies from other researchers, and the concept of efficacy proves the perception of utility.[6]

The Unified Theory of Acceptance and Use of Technology; UTAUT) is considered an evolution of the TAM model. Its main objective is determining the factors influencing people to adopt new technologies. The initial model is based on six dimensions: Performance Expectation, Effort Expectation, Social Influence, Facilitating Conditions, Behavioral Intention, and Use behavior [7].

The Unified Theory of Acceptance and Use of Technology (UTAUT) model is a framework used to understand and predict the adoption and use of technology. Four key constructs influence Behavioral Intention (BI) and Use Behavior (UB). The UTAUT model can be mathematically represented using a set of equations or weighted linear models that correspond to behavioral intention (BI) and use behavior (UB).

The following equation may represent Behavioral Intention (BI):

$$BI = \beta_1 \cdot PE + \beta_2 \cdot EE + \beta_3 \cdot SI + \beta_4 \cdot FC + \epsilon_1$$

Where:

- PE: Performance Expectancy
- EE: Effort Expectancy
- SI: Social Influence
- FC: Facilitating Conditions
- $\beta_1, \beta_2, \beta_3, \beta_4$ Coefficients representing the influence of each construct on Behavioral Intention
- $\epsilon_1$ Error term

Use Behavior (UB) may be represented by the following equation:

$$UB = \gamma_1 \cdot BI + \gamma_2 \cdot FC + \epsilon_2$$

Where:

- BI: Behavioral Intention
- FC: Facilitating Conditions
- $\gamma_1, \gamma_2$ : Coefficients representing the influence of Behavioral Intention and Facilitating Conditions on Use Behavior
- $\epsilon_2$: Error term

The impact of the key constructs is moderated by variables such as Gender (G), Age (A), Experience (E), and Voluntariness of Use (V). This introduces interaction terms, modifying the equations.

The following equation may now represent Behavioral Intention (BI):

$$BI = (\beta_1 + \mu_1 \cdot G + \mu_2 \cdot A) \cdot PE + (\beta_2 + \mu_3 \cdot E) \cdot EE + (\beta_3 + \mu_4 \cdot V) \cdot SI + \beta_4 \cdot FC + \epsilon_1$$

Use Behavior (UB) may now be represented by the following equation:

$$UB = (\gamma_1 + \mu_5 \cdot G + \mu_6 \cdot A) \cdot BI + (\gamma_2 + \mu_7 \cdot E) \cdot FC + \epsilon_2$$

Where:

- $\mu_1, \mu_2, \ldots \mu_7$: Coefficients for moderating effects of gender, age, experience, and voluntariness of use.

This mathematical formulation captures the linear relationships and moderating effects in the UTAUT model. These equations can be tested and calibrated through regression analysis, Structural Equation Modeling (SEM), or other statistical methods.

The Performance Expectation (ED) is associated with the individual's perception of the benefits that the use of technology will have on their professional performance and can be considered an evolution of the variable Perception of Utility of the TAM model. The variable Expectation of Effort (EE) is the degree of ease associated with using the system[7].

Social Influence (SI) is "the individual perception about the beliefs of others, regarding the use or not of this technology," while the Facilitating Conditions (FC) refer to the perception of the user regarding the existence of the resources necessary for the use of the technology in question.[8]

The second version of the model is called the Unified Theory of Acceptance and Use of Technology 2. The first changes were removing the moderating variable Voluntariness of Use and linking the variable Facilitating Conditions (FC) to the Behavior of Use (BU). New dimensions were included in UTAUT2, and the first, hedonic motivation, can be described as the degree of pleasure that the technology provides when used. The second dimension inserted in UTAUT2 was the Price Value in the acceptance and use of technology. Finally, the variables Habit and Experience were included, which affect the Behavior of Use (CU) and the Behavioral Intention (CI)[7].

The Motivation Model was developed to prove that people's behaviors are highly influenced by their motivations.[5]

The Theory of Planned Behavior is considered an evolution of the Theory of Rational Action, in which the only difference between the two theories is the presence of the dimension Perception of Behavioral Control [7]. This model has been used in several investigations, including studying individual acceptance and using new technologies.

The theory of DeLone and McLean is based on six dimensions: Information Quality, System Quality, Service Quality, System Use, Use Satisfaction, and Net [9] [10].

Information Quality refers to the quality of information the system can store or produce. Regarding the Quality of the System, the variables of effectiveness and efficiency are considered, affecting the benefits the client can extract from the quality of service, including the support provided by the developers or teams responsible. [10]

The variable System Use is related to the frequency of use of an information system and directly correlates with User Satisfaction: the greater the satisfaction, the greater the system use. The User Satisfaction dimension refers to customer satisfaction when using the information system and is directly related to the Net Benefits associated with the use of the system: greater satisfaction leads to greater use of the system and, consequently, more benefits.[9]

Finally, Net Benefits can be defined as the individual or organizational impacts that the system generates. This dimension is directly related to System Use and Use Satisfaction since the more benefits the system generates, the greater the user's satisfaction and use.[9]

Some changes were made in this model, such as the replacement of the word "benefits" with "impacts" since the expression "benefits" is associated with positive results, while the word "impacts" can already indicate positive or negative results.

System Quality was described as the result of measuring the convenience of access, flexibility, integration, and system response time.[11]

The information quality dimension may be defined as the perception of the importance and usefulness of information items. To evaluate the quality of the information, the variable's accuracy and relevance of the report must be taken into account.[11]

The Use of Information Systems was approached as one of the main concepts in the descriptive model of information systems in organizational contexts, and the use can be divided into three individual levels: level 1 refers to the use that leads to management actions, level 2 leads to the creation of significant changes and level 3 is distinguished as the recurrent use of a system. However, two years earlier, there were four levels of use, each with a different purpose – to obtain instructions, store data, carry out control actions, and carry out planning actions.[11]

Net Impacts were defined as Individual and Organizational Impacts [7]. The Individual Net Impacts can be analyzed according to the time spent performing tasks using the system. In order to realize the individual benefits of using a system, the concepts of accuracy of interpretation and quality of the decision must be analyzed. Finally, increased productivity is the main advantage of using an information system. The themes were organizational impacts, profits, improvement of results, and processes.

The opinion about the future and morality of the area of Artificial Intelligence does not generate consensus among the citizens of the different parts of the World: one part of the population believes that it will solve all the problems, while the other thinks that it will bring quite negative consequences for the World.[12]

Artificial intelligence varies between continents and countries since the policies adopted by their governments influence the lives of people and companies. In Europe, there are divergences in individual opinions, probably because the main laboratories of Artificial Intelligence are located outside this continent.[12]

Ethics is related to the ability to analyze a situation, perceive whether it is positive or negative, and act according to our ideology. However, the concept is not so simple to explain since, on most occasions, the dilemma is not related to simple positive or negative situations. In this way, Ethics is highly influenced by our experiences and life experiences, and it can be defined as a regulation of conduct that we create to define our actions and that usually ends up following legal and social ideologies and regulations.[12] [13]

For many years, due to factors associated with ethics, the creation and implementation of new technologies have generated discussion and fear in the general population, as happened with nuclear energy or cars. In most cases, laws and regulations are created, as has already happened with the General Data Protection Regulation, which protects people from the potentially harmful consequences that technology may cause [14].

The issue of Ethics in Artificial Intelligence has been debated with great frequency in recent years because the machines and algorithms that use this technology have as their main objective to make them as similar as possible to humans, either to make life easier for people, reduce the time spent on the accomplishment of a task, or even perform functions that an individual would not be able to.[15]

Many people raise ethical questions, as they believe that implementing Artificial Intelligence systems will create an exaggerated number of robots, loss of professions, and other consequences.[13]

Ethics can be used in different areas, and the definition of Ethics on the Internet can be replicated in Artificial Intelligence. Internet Ethics defines what is morally and ethically acceptable to do on the Internet since this is a very conducive medium for fraud and attacks on personal data. For this reason, the dimension Attitude Towards Ethical Behavior was created, which evaluates the probability of a person performing a positive or negative action when using the Internet, having emerged as an adaptation of the original dimension of the Theory of Planned Behavior. Performance Expectation is directly related to the Behavioral Intent of products that use Artificial Intelligence. [15]

The existence of infrastructures and conditions considered sufficiently solid to use a given technology has an impact on the intention and use of it, while Social Influence may contribute to a higher probability of use [16]

The way of thinking is a consequence of ethical values and individual beliefs, and the dimension Attitude Towards Ethical Behavior emerges as the application of this concept in the digital environment of the Internet. Finally, the Net Benefits variable covers the benefits associated with the Use of Artificial Intelligence, both at the personal and organizational level and is directly affected by the use of technologies.[13]

### III. RESEARCH MODEL

In order to answer the research question presented in the Introduction, a model will be elaborated based on the theoretical concepts addressed throughout the Literature Review and on the dimensions of the models analyzed.

**Table I – Constructs.**

| Dimension | Definition | Author |
|---|---|---|
| Performance Expectation (EP) | Perception regarding the gains and benefits obtained through the Use of Artificial Intelligence in organizational processes. | [7] |
| Facilitating Conditions (CF) | Belief of the user in which he believes that he has the resources and the conditions, at a material and organizational level, to use Artificial Intelligence mechanisms in his company. | [7] |
| Social Influence (SI) | Degree to which an individual realizes that others believe he should use the new system, in this case Artificial Intelligence. | [7] |
| Behavioral Intention (IC) | Level of interest, on the part of the user, in the Use of Artificial Intelligence mechanisms through the existence of all the necessary conditions | [7] |
| Usage (U) | Regularity of Use of a particular system, in this case Artificial Intelligence. | [10] |
| Attitude Toward Behavior Ethical (AFCE) | Individual vision about the fundamental principles associated with the Use of the Internet, applied in this case to Artificial Intelligence. | [17] |
| Net Benefits (BL) | Consequential impacts of the use of a certain technology, in this case Artificial Intelligence. | [9] |

There is a causal link between the Performance Expectation of products containing Artificial Intelligence and the Behavioral Intent of potential customers[18]. In order to corroborate their hypothesis, the Theory of Motivation argues that to provoke the Behavioral Intention of a given technology in a person, the benefits associated with the same – called Extrinsic Motivation – must be presented.[5]

H1. Performance Expectation directly affects the individual's Behavioral Intent.

The dimension Facilitating Conditions (FC) corresponds to the user's belief, in which he believes that he has the resources and conditions, at a material and organizational level, to use Artificial Intelligence mechanisms in his company). There is a direct relationship between the Facilitating Conditions with both Social Influence and Behavioral Intention. The Behavioral Intention of Artificial Intelligence and its use is directly affected by the existing Enabling Conditions. That is, there is a set of technical requirements, such as the skills and characteristics of employees associated with the existence of infrastructures to support this technology, necessary to ensure the Use of Artificial Intelligence [19]

H2a. The Facilitating Conditions positively impact the user's Behavioral Intent.

H2b. Enabling Conditions positively impact Social Influence.

Social Influence (SI) is defined as the perception of an individual according to which he perceives that other people believe that he should use a new system, and in the proposed model, the system is Artificial Intelligence. There is a direct relationship between Social Influence and Behavioral Intention and the Use of Artificial Intelligence.[7]

The Social Influence dimension emerges as the interpretation and fusion of the UTAUT model with the Rational Action Theory (TRA), used in the TAM2 model, the Theory of Planned Behavior (TPB), and the joining model of the same[18]. The greater the Social Influence of products that use Artificial Intelligence, the greater the Behavioral Intent of users in their products[18]. There is a relationship between the influence of superiors and the Use of Artificial Intelligence, which can be explained by the fear of falling behind direct market competitors [19].

H3a. Social influence positively affects behavioral intention.

H3b. Social Influence Positively Affects the Use of Artificial Intelligence.

The variable Behavioral Intention (CI), which has a direct relationship with the Usage dimension, translates the level of interest on the user's part in the Use of Artificial Intelligence mechanisms through the existence of all the necessary conditions[7].

The behavioral intention dimension appears in the theories TAM, UTAUT/UTAU2, and Rational Action Theory (ART), which refers to the subjective norm, which states that the behaviors performed by people are determined by their will and intention. Thus, the stronger the behavioral intention is, the stronger the effective use of artificial intelligence will be. [18]

H4. There is a consequential relationship between Behavioral Intention and the Use of Artificial Intelligence.

Personality and ethical values can influence people's beliefs and decisions, and the attitude to ethical behavior (AFCE) is related to the individual's view of the fundamental principles associated with Internet use [17].In this case, the variable focuses on ethical values from a perspective turned to artificial intelligence, which positively influences use.

H5. The Attitude Towards Ethical Behavior directly influences the Use of Artificial Intelligence.

System Use is defined as the regular use of certain systems[10]. In the proposed model, this variable will be linked to the Use of Artificial Intelligence mechanisms within organizations and correlates with the variable Net Benefits.Net Benefits, from the second version of the Delone & McLean model, representing the consequent impacts of using a particular technology[9]. This variable arose from the junction of the dimensions of Organizational Impact and Individual Impact, which comprises all benefits associated with Artificial Intelligence [14].

H6. There is a correlation between the Use of Artificial Intelligence mechanisms and the Net Benefits dimension.:

IV. METHOD

This approach is suggested by several researchers [20] and is considered in the context of behavioral research[21]. In order to collect as much data as possible and test the proposed model, the online questionnaire will consist of a set of questions with answers on the Likert scale with values from 1 to 7, where 1 means "Strongly Disagree" and 7 "Strongly Agree". This choice is due to the ease of sending and receiving questionnaires online and the possibility of using data analysis methods more effectively.

As the target audience of the online questionnaire, the research will focus on employees of companies, with emphasis on the area of Information Technology, since it is the group with the closest proximity to possible processes and tools of Artificial Intelligence.

The statistical data were taken from 237 responses to the questionnaire.

Regarding the respondents' gender, there is a total of 128 men, equivalent to 54.01% of the sample, compared to 45.99% of females. In order to analyze the age, the sample was separated into people over 31 years of age and people aged 31 years or younger, which resulted in a clear predominance in the first group, with 74.68% of the sample, and only 60 people were 31 years of age or younger.

Finally, analyzing the data related to Educational Qualifications, it can be concluded that the vast majority of people, with a share of 49.37%, have a Bachelor's degree, and the remaining 50.63% of people are divided into Secondary Education (12.66%), Master's or Postgraduate (36.71%) and PhD with only three people (1.27%).

The structural equation model (SEM) with partial least squares (PLS) was used to evaluate the variables' relationships. PLS is used to validate the causality of structural models, theoretically explained earlier. The tool used was SmartPLS 3.0. This approach was used to study performance expectations, facilitating conditions, social influence, attitude toward ethical behavior (independent variables), behavioral intention, use, and net benefits (dependent variable). PLS is suitable for this research study as we can use it in small samples with non-normal distribution. In addition, it decreases the residual variance of the dependent variables (Hair et al., 2017). Although the selected dimensions have already been used in previous investigations, the measurement model was tested to evaluate the reliability and validity of the dimensions. Thus, the measurement model was examined through different tests, such as dimension reliability, internal reliability, convergent validity, and discriminant validity.

The objective of this chapter is to perform a complete validation of the data obtained and the proposed model so that several analyses will be performed. Initially, a presentation of the sample and the descriptive data of the questionnaire, such as the age, gender, and educational qualifications of the respondents, will be made. The measurement model will be analyzed, and the reflective dimensions will be presented based on the criteria Composite Reliability, Cronbach's Alpha, Outer Loadings, Average Variance Extracted, Cross-Loading, and Fornell-Larcker. In section 4.3., the results of the structural model will be discussed, in which the values of the Inner VIF, Coefficient of Determination (V2), and F² will be analyzed.

## V. Analysis and Results

The evaluation of the model used in the initial phase focuses on the measurement model, and the reflective dimensions' validity and reliability levels should be analyzed.[22]

First, the internal consistency should be analyzed through the correlation between the data obtained from the answers, using Cronbach's Alpha. In order to meet this criterion, all values must be greater than 0.7, and the higher it is, the greater its consistency.

Then, to complement the evaluation, the Composite Reliability criterion should be used, similar to Cronbach's Alpha but which differs in terms of the valuation of the dimensions. The Composite Reliability considers that the dimensions have different weights in the model and that values greater than 0.6 must be accepted to guarantee reliability.[22]

Analyzing the values referring to the Composite Reliability and Cronbach's Alpha, present in Annex 4, it is verified, concerning the Composite Reliability, that all dimensions have values higher than 0.7, close to 1, ensuring excellent reliability. Concerning Cronbach's Alpha, except the Facilitating Conditions, all other dimensions have values considered good at the level of Cronbach's Alpha (greater than 0.8).

Convergent validity shall be obtained using the Outer Loadings and Average Variance Extracted (AVE) indicators, which are represented in Annex 4. The Average Variance Extracted – AVE – allows us to evaluate if the dimensions have convergent validity, that is, if the indicators belong to and explain the associated dimension, and the value of the same must be higher than 0.5 (Henseler et al., 2009). The Outer Loadings, or Reliability Indicators, demonstrate whether the items of a dimension are associated with each other, the ideal being higher values and with minimum values of acceptance 0.7 [23]

Thus, analyzing the values present in Annex 4, it is verified that in the Outer Loading column, all the values are higher than the acceptable value of 0.7, and the values of Average Variance Extracted are higher than 0.5, which indicates that there is Convergent Validity.

Discriminant validity refers to the differentiation of the constructs of the model, and they must be solely explained by themselves and not by others. That is, it evaluates the level of unmistakability of the constructs of the model. In order to perform the discriminant validity analysis, the Cross-Loadings and Fornell-Larcker criteria will be used.

The Cross-Loadings criterion should be considered fulfilled if each variable is superior in relation to the others, in terms of value, that is, if the item in question has a higher loading in its item compared to the other dimensions (Hair et al., 2017). As noted in Annex 2, this criterion is fully valid.

The Fornell-Larcker method, also used to measure discriminant validity, relates the square root of the Extracted Mean Variance (AVE) – first value – with the correlations of the latent variables – second value -, and in order to have a full confirmation of this criterion, the first value must be higher than the second (Ringle, 2014) . Once again, as can be seen in Table II, there was a full confirmation of the criterion in question, and all values corresponded to the requirement

**Table II Inner VIF**

|      | AFCE | BL | CF | EP | IS | IC | Use |
|------|------|----|----|----|----|----|-----|
| AFCE |      |    |    |    |    |    | 1,284 |
| BL   |      |    |    |    |    |    |     |
| CF   |      |    |    | 1,000 |  |    | 1,429 |
| EP   |      |    |    |    |    | 1,369 |     |
| IS   |      |    |    |    |    | 1,320 | 1,185 |
| IC   |      |    |    |    |    |    | 1,482 |
| Use  |      | 1,000 |  |    |    |    |     |

In order to analyze the collinearity, the Inner Variance Inflator Factor (Inner VIF) will be used, and for it to be confirmed, the values must be less than 5. All Inner VIF values vary between 0.768 and 0.921, which proves and demonstrates that there are no collinearity problems

**Table III Fornell–Larcker criterion**

|      | AFCE  | BL    | CF    | EP    | IS    | IC    | Use   |
|------|-------|-------|-------|-------|-------|-------|-------|
| AFCE | **0,811** |       |       |       |       |       |       |
| BL   | 0,517 | **0,846** |   |       |       |       |       |
| CF   | 0,320 | 0,544 | **0,768** |   |       |       |       |
| EP   | 0,425 | 0,767 | 0,474 | **0,885** |  |       |       |
| IS   | 0,158 | 0,452 | 0,442 | 0,400 | **0,914** |   |       |
| IC   | 0,470 | 0,758 | 0,558 | 0,739 | 0,394 | **0,921** |   |
| Use  | 0,539 | 0,781 | 0,441 | 0,712 | 0,325 | 0,805 | **0,890** |

The Coefficient of Determination is evaluated on a scale from 0 to 1, in which high values are equivalent to 0.75, mean values to 0.5, and weak values around 0.25. Analyzing Table IV, we can define the variables Net Benefits, Behavioral Intention, and Use as strong determination coefficients since they present values of 0.610, 0.603, and 0.681, respectively, and the variable Social Influence as weak, given that it presents a value of 0.195 (Hair et al., 2017).

Thus, through the Coefficient of Determination, it can be affirmed that the structural model can explain the latent variables Net Benefits (BL), Behavioral Intention (CI) and Use (U), although it cannot do so for the Social Influence dimension.

**Table IV $R^2$**

| CONSTRUCTS | $R^2$ |
|------------|-------|
| BL | 0,610 |
| IS | 0,195 |
| IC | 0,603 |
| U  | 0,681 |

The effect of the $F^2$ criterion can be determined according to different levels of values: a value greater than 0.350 has a large effect, while a value between 0.150 and 0.350 has an average effect, values between 0.02 and 0.150 reveal a small effect, and any value below 0.02 should be rejected. The values referring to $F^2$ are shown in Table V. In this case, the hypotheses H3a and H3b were immediately rejected since they presented values between 0.003 and 0.001. [23]

The variable Performance Expectation has a significant effect (0.667) on Behavioral Intention while Facilitating Conditions influence Behavioral Intention in a small way since the value of $F^2$ is only 0.114. On the other hand, Facilitating Conditions have an average effect on Social Influence ($F^2 = 0.243$), and Attitude towards Ethical Behavior, hypothesis H5, has a small effect on the Use of Artificial Intelligence. Finally, with values above 0.350 $F^2$ (1.041 and 1.561, respectively), Behavioral Intent affects use, and use affects Net Benefits

Table V - Results

| Hipothesys | Beta | PVal. | $F^2$ | Effect | Decision |
|---|---|---|---|---|---|
| H1. Performance Expectation directly affects the individual's Behavioral Intention. | 0.602 | 0 | 0.667 | Large | **sup** |
| H2a. Enabling Conditions positively impact the user's Behavioral Intent. | 0.225 | 0 | 0.114 | Small | **sup** |
| H2b. Enabling Conditions positively impact Social Influence. | 0.442 | 0 | 0.243 | Medium | **sup** |
| H3a. Social influence positively affects behavioral intention. | 0.040 | 0.453 | 0.003 | - | **Not sup** |
| H3b. Social Influence Positively Affects the Use of Artificial Intelligence. | 0.016 | 0.725 | 0.001 | - | **Not sup** |
| H4. There is a consequential relationship between Behavioral Intention and the Use of Artificial Intelligence. | 0.701 | 0 | 1.041 | Large | **sup** |
| H5. The Attitude Towards Ethical Behavior directly influences the Use of Artificial Intelligence- | 0.207 | 0 | 0.105 | Small | **Supported** |
| H6. There is a correlation between the use of artificial intelligence mechanisms and the dimension of net impacts. | 0.706 | 0 | 1.561 | Large | **Supported** |

Supp - Supported

Analyzing the hypotheses initially created in the proposed model, it will be assumed that a hypothesis becomes valid if the respective estimated coefficient is significant. That is, its value referring to the P-Value column – observed in table V – is less than 0.05. In this sense, of the total of the nine assumptions, only two were not verified in the data analysis.

As can be verified by the level of significance, the Performance Expectation directly affects the Behavioral Intention of the individuals, proving hypothesis 1. In fact, this result is corroborated by some researchers (e.g. [18]) since hypothesis 4 of his work studied the correlation between the variables of Performance Expectation and Behavioral Intention.

The opinion of individuals about the potential gains from adopting a technology, in this case Artificial Intelligence, directly affects their decision. Therefore, if someone considers that AI will bring advantages to their performance or the quality of their work, then there is a very high probability that they intend to use it.[7]

Hypotheses H2a and H2b were proven since both presented null values of P-Value, which means that it was confirmed that the Facilitating Conditions positively impact both Behavioral Intention and Social Influence. The hypotheses were elaborated with the assumption that if there are resources, knowledge, and a consensus on the part of those who surround and influence an individual to adopt Artificial Intelligence in their tasks, then it would be very likely that the same would want to do so. These results coincide with those published by [19], in which it is proven that the Enabling Conditions influence Behavioral Intention and the Use of Artificial Intelligence materialized in necessary skills and infrastructures.

It was not possible, however, to validate the hypotheses H3a and H3b, associated with the variable Social Influence, which defends the same positively and directly affects the Behavioral Intention and the Use of technology since the P-value values were higher than 0.05 (0.453 and 0.725, respectively). Studies by [18] and [19] have proven that social influence affects the intent and use of artificial intelligence. It should be noted, however, that in previous studies regarding the adoption of other types of systems, this hypothesis has not been validated [8]

The P-value values related to hypotheses H4 and H6, being lower than 0.05, corroborate the idea that Behavioral Intention affects the Use of Artificial Intelligence, which in turn presents a correlation with the Net Benefits associated with the technology in question. The greater the intention to adopt a technology, the more likely it is to be used by the individual and obtain benefits.

Hypothesis 4 was validated, corroborating the hypotheses that argued and proved that the higher the Behavioral Intent, the more likely the use of an Artificial Intelligence product.[16] [18]

Finally, the construct related to the Attitude Towards Ethical Behavior directly influences the Use of Artificial Intelligence (Hypothesis 5), as can be verified by the level of significance below the established level. It can be proven that the opinion and beliefs of an individual about the importance of the rules and ethical norms of this technology can contribute positively to the Use of Artificial Intelligence., which corroborates the idea that awareness of Ethics in Artificial Intelligence leads to its adoption [24].

In this way, it was possible to prove, based on the data obtained through the online questionnaire, that seven of the nine idealized research hypotheses were validated and supported by scientific models, including hypothesis 5, which defends the existence of a correlation between the Attitude Towards Ethical Behavior and the Use of Artificial Intelligence. The Social Influence dimension did not obtain values considered sufficient in the P-Value and F² fields, so it was rejected.

## VI. Conclusion

Over the last few years, we have witnessed exponential growth in artificial intelligence, mainly due to the daily data generated. The volume of data is increasing, which can be justified by people's network navigation, especially on social networks, and by the existence of more and more data sources, such as sensors, radars, and cameras. However, in recent times, people have become increasingly fearful about the security and privacy of their networked data and the appearance of robots in society, so the ethical issue of artificial intelligence has been called into question.

Then came the question of investigation of this work that seeks to understand how it is possible to adopt Artificial Intelligence in society, respecting and valuing the issue of Ethics of data and people. To this end, a model was developed consisting of seven dimensions, of which Ethics is one of the pillars, and seven research hypotheses, which were validated by the data obtained through a questionnaire, with a sample of 237 people, mostly from the area of Information Technology.

To evaluate the relationships of the variables, we used the structural equation model with partial least squares, also called Structural Equation Model partial least square (SEM-PLS), which resulted in the validation of seven of the nine research hypotheses outlined and the two rejected hypotheses were related to the correlation of the Social Influence dimension with the variables Behavioral Intention and Use of Artificial Intelligence.

According to the values obtained, it is possible to affirm that the Facilitating Conditions positively impact the Behavioral Intention (Hypothesis H2a) and the Social Influence (Hypothesis H2b) and that the Performance Expectation directly affects the Behavioral Intention of the Individual (Hypothesis 1). The consequential relationship between the Intention to Use and the Use of Artificial Intelligence (Hypothesis 4), and Hypothesis 6, which demonstrated that the Use of Artificial Intelligence leads to the achievement of Net Benefits, were also proved. Finally, the research question initially defined is answered mainly by Hypothesis 5, which means that there is evidence that demonstrates that ethics materialized in the dimension of attitude towards ethical behavior impacts the intention to use this technology. It is thus proven that Ethics affects people's decisions.

The theme of this work arose after realizing the importance of data protection and ethical issues in today's society, as well as the need to reconcile these two pillars with the growth of the Use of Artificial Intelligence. The model created in this work may become the starting point for new works that study the Use or adoption of Artificial Intelligence in society in more detail or serve as a reference for new investigations, and it has been proven the possibility of adopting this technology in an ethical way in today's society. It is also possible that the investigations carried out in this work, materialized in the final proven model, will be used at an individual or organizational level by companies that already use or want to adopt Artificial Intelligence in an Ethical way, ensuring that the protection and privacy of their customers are safeguarded.

For possible future work, I think it would be interesting to try to apply the model created to each specific area of Artificial Intelligence and see if the type of results obtained would vary with the increase of the sample used by the questionnaire. Another interesting idea would be to try to adapt this model to new emerging technologies that emerge in the coming years to understand if society's evolution would directly affect the importance given by people to the ethical values associated with technology.

Finally, I think it might be relevant to understand the importance of gender and nationality in highlighting the issue of Ethics in adopting Artificial Intelligence. The goal would be to analyze whether these two variables directly affect the use and adoption of artificial intelligence.

## References

[1] S. Aparicio, J. T. Aparicio, and C. J. Costa, "Data Science and AI: Trends Analysis," in *2019 14th Iberian Conference on Information Systems and Technologies (CISTI)*, Jun. 2019, pp. 1–6. doi: 10.23919/CISTI.2019.8760820.

[2] C. J. Costa and M. Aparicio, "Applications of Data Science and Artificial Intelligence," *Appl. Sci.*, vol. 13, no. 15, Art. no. 15, Jan. 2023, doi: 10.3390/app13159015.

[3] J. T. Aparicio, M. Romao, and C. J. Costa, "Predicting Bitcoin prices : The effect of interest rate, search on the internet, and energy prices," in *2022 17th Iberian Conference on Information Systems and Technologies (CISTI)*, Jun. 2022, pp. 1–5. doi: 10.23919/CISTI54924.2022.9820085.

[4] E. M. Rogers, "Diffusion of Innovations: Modifications of a Model for Telecommunications," in *Die Diffusion von Innovationen in der Telekommunikation*, M.-W. Stoetzer and A. Mahler, Eds., Berlin, Heidelberg: Springer, 1995, pp. 25–38. doi: 10.1007/978-3-642-79868-9_2.

[5] F. D. Davis, "Perceived Usefulness, Perceived Ease of Use, and User Acceptance of Information Technology," *MIS Q.*, vol. 13, no. 3, pp. 319–340, 1989, doi: 10.2307/249008.

[6] M. Aparicio and C. J. Costa, "Collaborative systems: characteristics and features," in *Proceedings of the 30th ACM international conference on Design of communication*, in SIGDOC '12. New York, NY, USA: Association for Computing Machinery, Oct. 2012, pp. 141–146. doi: 10.1145/2379057.2379087.

[7] V. Venkatesh, M. G. Morris, G. B. Davis, and F. D. Davis, "User Acceptance of Information Technology: Toward a Unified View," *MIS Q.*, vol. 27, no. 3, pp. 425–478, 2003, doi: 10.2307/30036540.

[8] I. Pedrosa, C. J. Costa, and M. Aparicio, "Determinants adoption of computer-assisted auditing tools (CAATs)," *Cogn. Technol. Work*, vol. 22, no. 3, pp. 565–583, Aug. 2020, doi: 10.1007/s10111-019-00581-4.

[9] William H. Delone and Ephraim R. McLean, "The DeLone and McLean Model of Information Systems Success: A Ten-Year Update," *J. Manag. Inf. Syst.*, vol. 19, no. 4, pp. 9–30, Apr. 2003, doi: 10.1080/07421222.2003.11045748.

[10] W. H. DeLone and E. R. McLean, "Information Systems Success: The Quest for the Dependent Variable," *Inf. Syst.*

*Res.*, vol. 3, no. 1, pp. 60–95, Mar. 1992, doi: 10.1287/isre.3.1.60.

[11] F. Bento, C. J. Costa, and M. Aparicio, "S.I. success models, 25 years of evolution," in *2017 12th Iberian Conference on Information Systems and Technologies (CISTI)*, Jun. 2017, pp. 1–6. doi: 10.23919/CISTI.2017.7975884.

[12] M. Piteira, M. Aparicio, and C. J. Costa, "Ethics of Artificial Intelligence: Challenges," in *2019 14th Iberian Conference on Information Systems and Technologies (CISTI)*, Jun. 2019, pp. 1–6. doi: 10.23919/CISTI.2019.8760826.

[13] C. J. Costa, M. Aparicio, S. Aparicio, and J. T. Aparicio, "The Democratization of Artificial Intelligence: Theoretical Framework," *Appl. Sci.*, vol. 14, no. 18, Art. no. 18, Jan. 2024, doi: 10.3390/app14188236.

[14] N. Urbach and B. Müller, "The Updated DeLone and McLean Model of Information Systems Success," in *Information Systems Theory: Explaining and Predicting Our Digital Society, Vol. 1*, Y. K. Dwivedi, M. R. Wade, and S. L. Schneberger, Eds., New York, NY: Springer, 2012, pp. 1–18. doi: 10.1007/978-1-4419-6108-2_1.

[15] P. Railton, "Ethical Learning, Natural and Artificial," in *Ethics of Artificial Intelligence*, S. M. Liao, Ed., Oxford University Press, 2020, p. 0. doi: 10.1093/oso/9780190905033.003.0002.

[16] V. Venkatesh, J. Y. L. Thong, and X. Xu, "Consumer Acceptance and Use of Information Technology: Extending the Unified Theory of Acceptance and Use of Technology," *MIS Q.*, vol. 36, no. 1, pp. 157–178, 2012, doi: 10.2307/41410412.

[17] Y.-Y. Wang, Y.-S. Wang, and Y.-M. Wang, "What drives students' Internet ethical behaviour: an integrated model of the theory of planned behaviour, personality, and Internet ethics education," *Behav. Inf. Technol.*, vol. 41, no. 3, pp. 588–610, Feb. 2022, doi: 10.1080/0144929X.2020.1829053.

[18] O. A. Gansser and C. S. Reich, "A new acceptance model for artificial intelligence with extensions to UTAUT2: An empirical study in three segments of application," *Technol. Soc.*, vol. 65, p. 101535, May 2021, doi: 10.1016/j.techsoc.2021.101535.

[19] G. Cao, Y. Duan, J. S. Edwards, and Y. K. Dwivedi, "Understanding managers' attitudes and behavioral intentions towards using artificial intelligence for organizational decision-making," *Technovation*, vol. 106, p. 102312, Aug. 2021, doi: 10.1016/j.technovation.2021.102312.

[20] C. J. Costa, J. T. Aparicio, and M. Aparicio, "Socio-Economic Consequences of Generative AI: A Review of Methodological Approaches," Nov. 14, 2024, *arXiv*: arXiv:2411.09313. doi: 10.48550/arXiv.2411.09313.

[21] J. T. Aparicio, M. Aparicio, and C. J. Costa, "Design Science in Information Systems and Computing," in *Proceedings of International Conference on Information Technology and Applications*, S. Anwar, A. Ullah, Á. Rocha, and M. J. Sousa, Eds., Singapore: Springer Nature, 2023, pp. 409–419. doi: 10.1007/978-981-19-9331-2_35.

[22] C. J. Costa, E. Ferreira, F. Bento, and M. Aparicio, "Enterprise resource planning adoption and satisfaction determinants," *Comput. Hum. Behav.*, vol. 63, pp. 659–671, Oct. 2016, doi: 10.1016/j.chb.2016.05.090.

[23] M. Aparicio, C. J. Costa, and R. Moises, "Gamification and reputation: key determinants of e-commerce usage and repurchase intention," *Heliyon*, vol. 7, no. 3, Mar. 2021, doi: 10.1016/j.heliyon.2021.e06383.

[24] L. N. K. Leonard, T. P. Cronan, and J. Kreie, "What influences IT ethical behavior intentions—planned behavior, reasoned action, perceived importance, or individual characteristics?," *Inf. Manage.*, vol. 42, no. 1, pp. 143–158, Dec. 2004, doi: 10.1016/j.im.2003.12.008.